\documentclass[%
 aip,
 jmp,%
 amsmath,amssymb,
reprint,%
]{revtex4-1}

\usepackage{graphicx}
\usepackage{dcolumn}
\usepackage{bm}

\begin{document}

\title{Geometric spin manipulation  in semiconductor quantum dots }

\author{Sanjay Prabhakar}
\email[]{sprabhakar@wlu.ca}
\homepage[]{http://www.m2netlab.wlu.ca}
\affiliation{M\,$^2$NeT Laboratory, Wilfrid Laurier University, 75 University Avenue West, Waterloo, ON, Canada, N2L 3C5
}
\author{Roderick Melnik}
\affiliation{M\,$^2$NeT Laboratory, Wilfrid Laurier University, 75 University Avenue West, Waterloo, ON, Canada, N2L 3C5
}
\author{Akira Inomata}
\affiliation{
Department of Physics, State University of New York at Albany, New York 12222 USA
}

\date{March 28, 2014}

\begin{abstract}
We propose a  method to flip the spin completely by  an  adiabatic transport of quantum dots.
We show that it is  possible to flip the spin by inducing a geometric phase   on the spin state of a quantum dot. We estimate  the geometric spin flip time (approximately 2 pico-sec) which turned out to be much shorter than the experimentally  reported decoherence time (approx. 100 nano-sec) that would provide an alternative means of fliping the spin before reaching decoherence. It is important that both the Rashba coupling and the Dresselhaus coupling are present for inducing a phase necessary for spin flip.
If one of them is absent, the induced phase is trivial and irrelevant for spin-flip.
\end{abstract}

\maketitle

Manipulation of single electron spins  in low dimensional semiconductor nanostructures such as quantum dots (QDs),  wells and nanowires is essential for spintronic and spin-based quantum information processing.~\cite{nowack07,pechal12,prabhakar09,pryor06}
The electron spin in these nanostructures can be manipulated by several different techniques: for example, electron spin resonance induced by oscillating magnetic fields at the Zeeman frequency, and  electric field control of spin through spin-orbit coupling.~\cite{koppens06,amasha08,prabhakar13} More robust techniques have been suggested for manipulation of the electron spin in a QD by letting the dot to move adiabatically along a closed loop.~\cite{jose08,prabhakar10,ban12,jose06} Recently the geometric phase has been measured experimentally for qubits  driven by a microwave pulse in the presence of  tilted magnetic fields.~\cite{leek07,pechal12}

In this paper, we let a QD move adiabatically in the two dimensional (2D) plane of the electron gas under application of a gate controlled periodic lateral electric field. We report that in the presence of both the Rashba and the Dresselhaus spin-orbit couplings, the geometric phase changes its sign over a short period of time on the spin state. This tells us that the complete spin flip is possible by controlling the applied electric field. We point out that the geometric spin flip can be achieved much faster than the decoherence.

We write the total Hamiltonian of  a QD formed in the  plane of a two dimensional electron gas with gate potential confined along z-direction at the heterojunction in the  presence of externally  applied magnetic field along z-direction  as: $H = H_{0}  +  H_R+ H_D$.~\cite{prabhakar09,prabhakar11,golovach06} Here
\begin{eqnarray}
H_{0} = {\frac {\left\{\mathbf{p}+e\mathbf{A}(\mathbf{r})\right\}^2}{2m}} + {\frac{1}{2}} m \omega_o^2\mathbf{r}^2 + e\mathbf{E}(t)\cdot \mathbf{r} + \frac{\Delta}{2} \sigma_z,\label{hxy}\\
H_R =\frac{\alpha}{\hbar}\left\{\sigma_x \left(p_y+eA_y\right) - \sigma_y \left(p_x+eA_x\right)\right\},\label{rashba}\\
H_D= \frac{\beta}{\hbar}\left\{-\sigma_x \left(p_x+eA_x\right) + \sigma_y \left(p_y+eA_y\right)\right\}. \label{dresselhaus}
\end{eqnarray}
In~(\ref{hxy}), $\mathbf{p} = -i\hbar (\partial_x,\partial_y,0)$ is  the canonical momentum, $\mathbf{r}=\left(x,y,0\right)$  the position vector,    $e $ the electronic charge, $m$ the effective mass of an electron in the QDs,  $\Delta=g_0\mu_B B$  the Zeeman energy,  $\mu_B$ is  the Bohr magneton and $g_0$  the bulk g-factor of an electron in the  QD. Also, $\mathbf{E}(t)=\left(E_x(t),E_y(t),0\right)$ with $E_x(t)=E_0\cos\omega t$ and $E_y(t)=E_0\sin\omega t$ is the electric potential energy due to the applied periodic lateral electric field. Note that $e\mathbf{E}(t)\cdot \mathbf{r}$ is the coupling energy (potential energy) having the dimension of energy. By varying $\mathbf{E}(t)$ very slowly,  we treat its two components as adiabatic parameters. The time varying electric field $\mathbf{E}(t)$ also induces a  magnetic field  which is in practice several order magnitude smaller than  the applied magnetic field along z-direction.~\cite{golovach06}   Thus, ignoring such a small contribution, we use the vector potential of the form  $\mathbf{A}(r)=B/2\left(-y,x,0\right)$. In~(\ref{rashba}),  $\alpha=\gamma_ReE_z$ is the Rashba spin-orbit coupling coefficient originating from structural inversion asymmetry. In~(\ref{dresselhaus}), $\beta=0.78\gamma_D\left(2me/\hbar^2\right)^{2/3}E_z^{2/3}$ is the Dresselhaus spin-orbit coupling coefficient originating from bulk inversion asymmetry. For GaAs QD, we chose $\gamma_R=4.4~\mathrm{{\AA}^2}$ and $\gamma_D=26~\mathrm{eV.{\AA}^3}$.

At a fixed time $t_0$, the electric field shifts   the center of QDs from $\mathbf{r}=0$ to $\mathbf{r}=\mathbf{r_0} \left(t_0\right)$, where $\mathbf{r}_0=-e\mathbf{E}\left(t_0\right)/m\omega_0^2$. Hence we write Hamiltonian~(\ref{hxy})  as:
\begin{equation}
H_{0} = {\frac {\left\{\mathbf{p}+e\mathbf{A}(\mathbf{r})\right\}^2}{2m}} + {\frac{1}{2}} m \omega_o^2\left(\mathbf{r}-\mathbf{r_0}\right)^2 -G + {\frac \Delta 2}\sigma_z,
\label{hxy-1}
\end{equation}
where  $G=r_0eE_0/2$ is an unimportant constant. As the applied $\mathbf{E}-$field varies, the QD will be adiabatically transported along a circle of radius $r_0=|\mathbf{r}_0|=eE_0/m\omega^2_0$.

Now, we write  relative coordinate $\mathbf{R}=\mathbf{r}-\mathbf{r_0}$ and   relative momentum $\mathbf{P}=\mathbf{p}-\mathbf{p_0}$, where $\mathbf{p_0}$ is the momentum of the slowly moving dot which may be classically given by $m\mathbf{\dot{r}}_0$.  We can show that the adiabatic variables $\mathbf{p}_0$ and $\mathbf{r}_0$ will be gauged away from the Hamiltonian by the transformation $\tilde{H}=UHU^{-1}$ and $\tilde{\psi}=U\psi$ with $U=\exp\left\{ \left(i/\hbar\right) \left(\mathbf{p_0}+e\mathbf{A}\left(\mathbf{r}_0\right)\right)\cdot \mathbf{R}\right\}$, so that
\begin{eqnarray}
\tilde{H}_{0} = {\frac {1}{2m}}\left\{\mathbf{P}+e\mathbf{A}(\mathbf{R})\right\}^2 + {\frac{1}{2}} m \omega_o^2R^2 -G+ {\frac \Delta 2}  \sigma_z,\label{tilde-hxy}\\
\tilde{H}_{so} = \tilde{H}_{R} +  \tilde{H}_{D} = UH_{so}\left(\mathbf{p},\mathbf{r}\right)U^{-1}=H_{so}\left(\mathbf{P},\mathbf{R}\right),\label{Hso}
\end{eqnarray}
where $\mathbf{A}\left(\mathbf{R}\right)=\left(B/2\right)\left(-Y,X,0\right)$. This means  the electron in the shifted dot obeys a quasi-static eigenequation, $\tilde{H}\left(\mathbf{P},\mathbf{R}\right)\tilde{\psi}_n\left(\mathbf{R}\right)=
\tilde{\varepsilon}_n\tilde{\psi}_n\left(\mathbf{R}\right)$, where $\tilde{H}=\tilde{H}_0+\tilde{H}_{so}$. By an adiabatic transport of the dot, the eigenfunction $\tilde{\psi}_n$ will acquire the geometric phase as well as the usual dynamical phase. Namely, $\psi_n\left(\mathbf{r},t\right)=e^{i\gamma_n\left(t\right)} e^{i\theta_n\left(t\right)}U^{-1}\tilde{\psi}_n\left(R\right)$, where $\gamma_n$ is the geometric phase and $\theta_n$ is the dynamical phase.

\begin{figure}
\includegraphics[width=6.8cm,height=6cm]{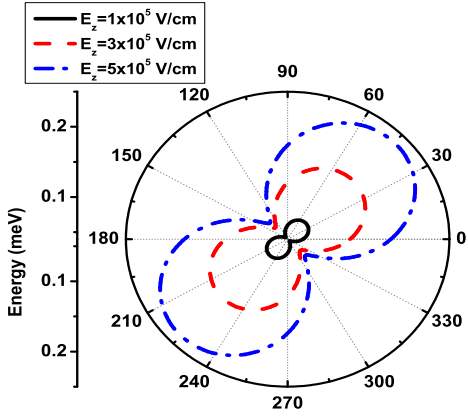}
\caption{\label{fig1}  Adiabatic control of eigenenergy  $\Delta \lambda=\lambda_+-\lambda_-$ (i.e., the energy difference between spin up  and down states) vs rotation angle. Here we chose $E_0=10^3~\mathrm{V/cm}$, $\ell_0=20~\mathrm{nm}$, $B=1~\mathrm{T}$.   }
\end{figure}

In order to evaluate the geometric phase explicitly, we write the original Hamiltonian $H$ in the form:
\begin{equation}
H=\tilde{H}_0\left(\mathbf{P},\mathbf{R}\right)+H_{so}\left(\mathbf{P},\mathbf{R}\right)+
H_{ad}\left(\mathbf{P},\mathbf{R};\mathbf{p_0},\mathbf{r_0}\right),
\label{total-adiabatic}
\end{equation}
where
\begin{eqnarray}
H_{ad}\left(\mathbf{P},\mathbf{R};\mathbf{p}_0,\mathbf{r}_0\right)&=&\frac{1}{m}\left\{\mathbf{P}
+e\mathbf{A}\left(\mathbf{R}\right)\right\}\cdot \left\{\mathbf{P}+e\mathbf{A}\left(\mathbf{R}\right)\right\}\nonumber\\
&&+H_{so}\left(\mathbf{p}_0,\mathbf{r}_0\right)+G',\label{Had}
\end{eqnarray}
with another unimportant constant $G'=r_0^2\omega_+^2/4$, where $\omega_+=\omega+\omega_c/2$ and  $\omega_c=eB/m$  is the cyclotron frequency.

The quasi-static Hamiltonian $\tilde{H}_0\left(\mathbf{P},\mathbf{R}\right)$ can be diagonalized on the basis of the number states $|n_+,n_-,\pm 1\rangle$:
\begin{equation}
\tilde{H}_0=\left(N_+ + \frac{1}{2}\right)\hbar\Omega_+ + \left(N_- + \frac{1}{2}\right)\hbar\Omega_- -G + \frac{\Delta}{2}\sigma_z,\label{H0}
\end{equation}
with $N_{\pm}=a^\dagger_{\pm}a_{\pm}$ are the number operators with eigenvalues $n_\pm \in N_0$. The other terms in~(\ref{Had}) can also be expressed in terms of the raising and lowering operators,
\begin{eqnarray}
H_{so}\left(\mathbf{P},\mathbf{R}\right)&=&\alpha\left(\xi_+\sigma_+ a_+ - \xi_- \sigma_- a_-\right)~~~~~~\nonumber\\
&&+i\beta\left(\xi_+\sigma_- a_+ + \xi_- \sigma_+ a_-\right)+H.c.,~~~~~~\label{Hso-1}\\
H_{ad}\left(\mathbf{P},\mathbf{R};\mathbf{p_0},\mathbf{r_0}\right)&=&\frac{\hbar}{2}\left(\xi_+z_+a_+-\xi_-z_-a_-\right)\omega_+ \nonumber\\
&&+\frac{1}{\hbar}\left(\alpha z_--i\beta z_+\right)m\omega_+\sigma_+ +H.c. \label{had}
\end{eqnarray}
In the above, we have used the notations, $z_{\pm}=x_0\pm iy_0$, $\xi_{\pm}= \sqrt {m\Omega/\hbar}\pm eB/\sqrt{4m\hbar\Omega}$, $\sigma_{\pm}=\left(\sigma_x\pm i\sigma_y\right) /2$,  $\Omega_{\pm}=\Omega \pm \omega_c/2$ and $\Omega=\sqrt {\omega_0^2+\omega_c^2/4}$. In~(\ref{Hso-1}) and~(\ref{had}), $H.c.$ signifies the Hermitian conjugate.
\begin{figure}
\includegraphics[width=8.5cm,height=7cm]{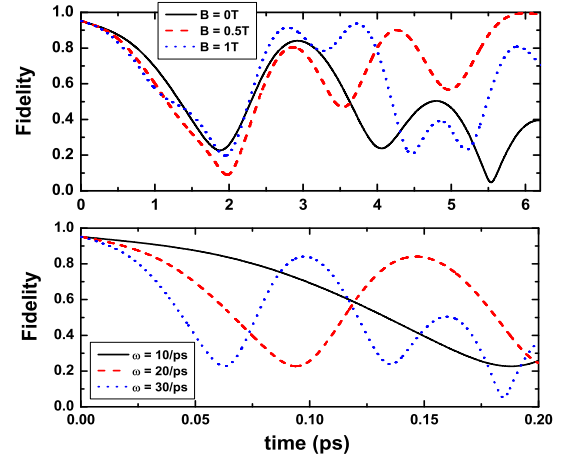}
\caption{\label{fig2} (Color online) Fidelity $=|\langle \psi_-\left(\theta\right)|\psi\left(\theta\right)\rangle|$ vs time.  For the case with no magnetic fields (solid line, upper panel), the spin Hamiltonian reduces to those of Ref.~\onlinecite{prabhakar10}. In the lower panel (B=0.01T),  much faster spin flip is achieved  with increasing adiabatic control frequency $\omega < \lambda_{\pm}/\hbar$. Here we chose $E_0=5\times10^4~\mathrm{V/cm}$, $\ell_0=20~\mathrm{nm}$ and $E_z=  10^5~\mathrm{V/cm}$.  }
\end{figure}
\begin{figure}
\includegraphics[width=8.5cm,height=4cm]{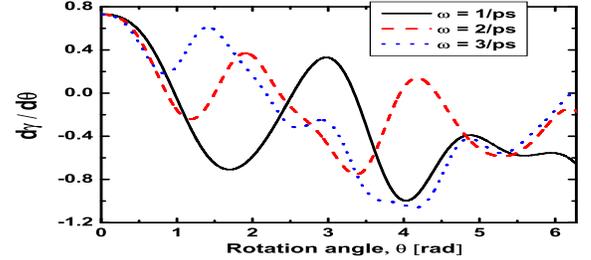}
\caption{\label{fig3} (Color online) Geometric phase ($d\gamma/d\theta$) vs rotation angle on the superposed state $|\psi\left(\theta\right)\rangle$. We see that the geometric phase changes from positive to negative regime. This tells us that the complete spin flip occurs only on the superposed state
during the adiabatic transport of the dots. Here we chose $E_0=10^3~\mathrm{V/cm}$, $\ell_0=20~\mathrm{nm}$, $B=1~\mathrm{T}$ and $E_z= 10^5~\mathrm{V/cm}$.    }
\end{figure}

To investigate the geometric phase during the adiabatic movement of the dots in the 2D plane, we write the quasi-spin Hamiltonian as:
\begin{equation}
h_{ad}=\kappa\left(Z_1-iZ_2\right)s_++\kappa\left(Z_1+iZ_2\right)s_- + \Delta s_z,\label{had-1}
\end{equation}
where $\kappa=m\omega_+/\hbar$,  $Z_1=\alpha x_0 + \beta y_0$,  $Z_2=\beta x_0 + \alpha y_0$ and  $s_{\pm}=s_x\pm  i s_y$ with  $s_x=\sigma_x/2,$ $s_y=\sigma_y/2$ and $s_z=\sigma_z/2$.    This is the only part of the complete Hamiltonian that contributes to the geometric phase. The eigenvalues of~(\ref{had-1}) are $ \lambda_{\pm} = \pm \kappa F$, where $F=\left(Z^2 + Z_1^2+Z_2^2\right)^{1/2}$ with $Z=\Delta/2\kappa$. In Fig.~\ref{fig1}, we have plotted the energy difference between $\lambda_+$ and  $\lambda_-$ vs rotation angle.  Modulation in the energy spectrum can be seen due to the fact that the energy spectrum of the dots depends on the adiabatic control parameters $x_0$ and $y_0$.

We construct a normalized orthogonal set of eigenspinors of Hamiltonian~(\ref{had-1}) as:
\begin{eqnarray}
\mathbf{\psi_+}\left(\theta\right)=\frac{1}{\sqrt {2F\left(Z+F\right)} }
\left(\begin{array}{c}
Z+F \\
Z_1+iZ_2\\
\end{array}\right),\label{chi-p-1}\\
\mathbf{\psi_-}\left(\theta\right)=\frac{1}{\sqrt  {2F\left(Z+F\right)} }
\left(\begin{array}{c}
Z_1-iZ_2\\
-Z-F\\
\end{array}\right).\label{chi-m-1}
\end{eqnarray}
Let us now calculate the geometric phase
\begin{widetext}
\begin{equation}
\gamma_+\left(\theta\right)=i \int_\theta \langle \psi_+ |  \partial_\theta | \psi_+\rangle d\theta= i \int_\theta \left[\sqrt{\frac{Z+F}{2F}} \partial_\theta \left\{ \sqrt{\frac{Z+F}{2F}} \right\} + \frac{Z_1-iZ_2}{\sqrt{2F\left(Z+F\right)}} \partial_\theta \left\{ \frac{Z_1-iZ_2}{\sqrt{2F\left(Z+F\right)}} \right\}\right] d\theta.\label{gamma-p}
\end{equation}
\end{widetext}
To find the exact analytical expression for the geometric phase of~(\ref{gamma-p}), we write $Z_1$ and $Z_2$ in a more convenient form:
\begin{equation}
Z_1= D \cos\left(\theta-\delta\right), ~ Z_2= D \sin\left(\theta+\delta\right),\label{z1}
\end{equation}
where $D=-r_0 \left( \alpha^2 + \beta^2 \right)^{1/2}$ and  $\tan \delta = \beta/\alpha$. By recognizing $\sin 2\delta=2\alpha\beta/\left(\alpha^2+\beta^2\right)$ and
\begin{equation}
F^2=Z^2 + D^2\left\{ 1+\frac{2 \alpha \beta}{\alpha^2+\beta^2}\sin\left(2\theta\right)   \right\},
\end{equation}
we write the expression for the geometric phase~(\ref{gamma-p}) as:
\begin{widetext}
\begin{equation}
\gamma_+\left(\theta\right)=-\frac{r_0^2}{2}|\alpha^2-\beta^2| \int_0^\theta \frac{1}{ \sqrt{ Z^2+D^2\left(1+\gamma_{_{RD}}\sin 2\theta \right)} \left\{Z+ \sqrt{   Z^2+D^2\left(1+\gamma_{_{RD}}\sin 2\theta \right)  } \right\} } d\theta,\label{gamma-p-1}
\end{equation}
\end{widetext}
where
\begin{equation}
\gamma_{_{RD}} = \frac{2\alpha\beta}{\alpha^2+\beta^2},~\cos\left(2\delta\right)=\frac{\alpha^2-\beta^2}{\alpha^2+\beta^2}.
\end{equation}
We clearly see that the Rashba-Dresselhaus spin-orbit coupling coefficients $\gamma_{_{RD}}$ couple to the adiabatic control parameters $E_x$ and $E_y$. Thus it is only possible to modulate the geometric phase on spin up and down spinors for the case of mixed  Rashba-Dresselhaus spin-orbit couplings.
The integral of~(\ref{gamma-p-1}) does not have a closed  form. However for the special cases such as  (i)  for the pure Rashba case ($\beta=0$), (ii) for  the pure Dresselhaus case ($\alpha=0$), and (iii)  for  $Z=0$,  the exact geometric phase as a function of rotation angle ($\theta$) is given by

\emph{(i) for the  pure Rashba case: }
\begin{equation}
\gamma_+\left(\theta\right)=-\frac{r_0^2 \alpha^2 \theta}{ 2\sqrt{ Z^2+D^2} \left\{Z+ \sqrt{   Z^2+D^2  } \right\} }.\label{gamma-p-2}
\end{equation}
\emph{(ii) for the  pure Dresselhaus  case: }
\begin{equation}
\gamma_+\left(\theta\right)=-\frac{r_0^2 \beta^2 \theta}{ 2\sqrt{ Z^2+D^2} \left\{Z+ \sqrt{   Z^2+D^2  } \right\} },\label{gamma-p-3}
\end{equation}
\emph{(iii) for  $Z=0$ i.e., without Zeeman energy:}
\begin{equation}
\gamma_+\left(\theta\right)=\frac{1}{2}\left[\tan^{-1}\left\{ \frac{\alpha^2+\beta^2}{|\alpha^2-\beta^2|}  \tan\theta  +\frac{2\alpha\beta}{|\alpha^2-\beta^2|}   \right\}\right]_0^\theta.\label{gamma-p-4}
\end{equation}
It is clear that for both the pure Rashba case (\ref{gamma-p-2}) and the pure Dresselhaus case (\ref{gamma-p-3}) the geometric phase is linearly proportional to the adiabatic parameter $\theta$ resulting only in Berry phases which induce no spin flip. In calculating the Berry phase for $Z=0$ case (\ref{gamma-p-4}), care must be taken (see Ref.~\onlinecite{footnote}).
It is more appropriate to divide the square-bracket into three portions as:
\begin{equation}
\left[  \mbox{\boldmath${\cdot}$}  \right]_0^\pi = \lim_{\epsilon \to 0} \left( \left[  \mbox{\boldmath${\cdot}$}  \right]_0^{\pi/2-\epsilon} +  \left[  \mbox{\boldmath${\cdot}$}  \right]_{\pi/2+\epsilon}^{3\pi/2-\epsilon} + \left[  \mbox{\boldmath${\cdot}$}  \right]_{3\pi/2+\epsilon}^{2\pi}\right).
\end{equation}
Notice that $\tan\left(\pi/2 - \epsilon\right) > 0$ whereas $\tan\left(\pi/2 + \epsilon\right) < 0$. As a result of this calculation, we can get the non-vanishing Berry phase:
\begin{equation}
\gamma_+\left(\theta\right)=\pi.\label{BP}
\end{equation}

To find the total geometric phase on the superposed states of $|\psi_+\rangle$ and $|\psi_-\rangle$, the evolution operator of~(\ref{had-1}) is needed.  Following Refs.~(\onlinecite{prabhakar10}) and~(\onlinecite{popov07}),   the exact evolution operator of~(\ref{had-1}) for a spin-1/2 particle can be written as:
\begin{equation}
U(t)=\left(\begin{array}{cc}\exp\left\{{\frac{b}{2}}\right\} + ac\exp\left\{{-\frac{b}{2}}\right\} & ~~~a\exp\left\{-{\frac{b}{2}}\right\}\\ \nonumber
c\exp\left\{-{\frac{b}{2}}\right\} & ~~~\exp\left\{-{\frac{b}{2}}\right\} \end{array}\right). \label{U}
\end{equation}
The  functions $a(t)$, $b(t)$, and $c(t)$ are given by
\begin{eqnarray}
\frac{d a}{d t}= \frac{k}{i\hbar}\left\{\left(Z_1-iZ_2\right)
-  a^2 \left(Z_1+iZ_2\right)  +2Z a \right\}, ~~~~~ \label{a}\\
\frac{d b}{d t}= \frac{2k}{i\hbar} \left\{ Z - \left( Z_1+iZ_2 \right) a \right\} ,~~~~~~~ \label{b}\\
\frac{d c}{d t}= \frac{k}{i\hbar} \left(Z_1+iZ_2\right) e^b. ~~~~~~\label{c}
\end{eqnarray}
At $\theta=0$, we use the initial condition
\begin{eqnarray}
\mathbf{\psi}\left(0\right)=\frac{1}{\sqrt 2 }
\left(\begin{array}{c}
\left\{ \frac{Z+\sqrt{Z^2+D^2}} {\sqrt{Z^2+D^2}}  \right\}^{1/2}\\
\frac{r_0\left(\alpha+i\beta\right)}{\left\{ \sqrt{Z^2+D^2}\left( Z+ \sqrt{Z^2+D^2} \right) \right\}^{1/2}}\\
\end{array}\right),\label{chi-0}
\end{eqnarray}
and write $\psi\left(\theta\right)=U(t,0)\psi\left(0\right)$ as
\begin{widetext}
\begin{equation}
\psi\left(\theta\right)=\frac{1}{\sqrt 2}\left(\begin{array}{c} \left[\exp\{b/2\} + a c \exp\{-b/2\}\right]\left\{ \frac{Z+\sqrt{Z^2+D^2}} {\sqrt{Z^2+D^2}}  \right\}^{1/2} +
a \exp\{-b/2\} \frac{r_0\left(\alpha+i\beta\right)}{\left\{ \sqrt{Z^2+D^2}\left( Z+ \sqrt{Z^2+D^2} \right) \right\}^{1/2}}  \\ \nonumber
c \exp\{-b/2\} \left\{ \frac{Z+\sqrt{Z^2+D^2}} {\sqrt{Z^2+D^2}}  \right\}^{1/2} +    \exp\{-b/2\} \frac{r_0\left(\alpha+i\beta\right)}{\left\{ \sqrt{Z^2+D^2}\left( Z+ \sqrt{Z^2+D^2} \right) \right\}^{1/2}} \end{array}\right). \label{chi}
\end{equation}
\end{widetext}
In Fig.~\ref{fig2}, we see that the fidelity is enhanced with magnetic field which means that the spin flip is much faster than for the case with no magnetic field during the adiabatic transport of the dots in the 2D plane. We also notice that the geometric spin flip time (less than 2 ps) is much faster than the experimentally reported decohenrence time ($\sim$100 ns)~\cite{nowack07} that might overcome the main danger for the low-temperature spin manipulation in QDs coming from the hyperfine coupling to the nuclear spins as earlier reported by Ban et. al in Ref.~\onlinecite{ban12}.

In Fig.~\ref{fig3}, we have plotted the geometric phase $d\gamma/d\theta=i\langle \psi \left(\theta\right) |\partial_\theta \psi\left(\theta\right) \rangle$ as a function of the rotation angle. As can be seen, the geometric phase can change from positive to negative values which tells us that the spin is completely  flipped   much faster than the decoherence time during the adiabatic transport of the dots.

To conclude, we have provided an alternative approach to flip the spin completely via the geometric phase. Our designed oscillating electric pulse is the smooth functions of $\sin\theta$ and $\cos\theta$ that might provide the simplest  way to measure the geometric phase in QDs. For the  experimental set up in the laboratory, see Ref.~\onlinecite{nowack07,pechal12}.   In Fig.~\ref{fig2}, we have shown that the fidelity is enhanced with increasing magnetic fields and frequency of the control pulse. The spin flip time (less than 2 ps)  is much faster than the experimentally reported decoherence time ($\approx 100 ns$). In Fig.~\ref{fig3}, we have shown that the sign change in the geometric phase indicates that the  spin is completely flipped during the adiabatic transport  of the dots. This result, yet to be experimentally verified, may provide an alternative for the manipulation of spins before reaching to decoherence. Either for the pure Rashba or the pure Dresselhaus spin-orbit coupling cases, the geometric phase on spin up and down spinors is linear in the rotation angle (see Eqs.~\ref{gamma-p-2} and \ref{gamma-p-3})  which is  not be useful if we are interested in flipping the spin.

This work was supported by NSERC and CRC programs, Canada. The authors acknowledge the Shared Hierarchical Academic Research Computing Network (SHARCNET) community and Dr. P.J. Douglas Roberts for his assistance and technical support.


\begin{thebibliography}{16}
\bibitem [{foo()}]{footnote}%
  \BibitemOpen
  \href@noop {} {}\bibinfo {note} {The quantity inside the square bracket on
  the right hand side of Eq.(22) vanish if one directly let
  $\theta=2\pi$.}\BibitemShut {Stop}%
\end{thebibliography}

%


\end{document}